\documentclass[conference,letter]{IEEEtran}
\usepackage{subcaption}     
\usepackage{makecell}
\usepackage{graphicx}
\usepackage{latexsym}
\usepackage{amsxtra} 
\usepackage{amsmath}
\usepackage{amsfonts}
\usepackage{marvosym}
\usepackage{scalefnt}
\usepackage{color,tabularx}
\usepackage{makecell}
\usepackage{booktabs,multicol,multirow}
\usepackage{algorithm}
\usepackage{comment}
\usepackage{url}
\usepackage{soul}
\usepackage{cite}
\usepackage{xcolor}
\usepackage[noend]{algpseudocode}
\usepackage[hidelinks]{hyperref}

\usepackage{fancyhdr}
\fancypagestyle{firstpage}{
  \fancyhf{}
  \fancyhead[C]{ 	To appear at the 2025 Design, Automation \& Test in Europe Conference \& Exhibition (DATE).}
 
}

\begin{document}\bstctlcite{IEEEexample:BSTcontrol}
\title{
Late Breaking Result: FPGA-Based Emulation and Fault Injection for CNN Inference Accelerators\vspace{-0.5em}
}

\author{\IEEEauthorblockN{Filip Masar, Vojtech Mrazek, Lukas Sekanina}
\IEEEauthorblockA{Brno University of Technology, Faculty of Information Technology, Brno, Czechia\\
Email: xmasar18@stud.fit.vutbr.cz, mrazek@fit.vutbr.cz, sekanina@fit.vutbr.cz\vspace{-1.8em}}}


\maketitle
\thispagestyle{firstpage}

\begin{abstract}
A new field programmable gate array (FPGA)-based emulation platform is proposed to accelerate fault tolerance analysis of inference accelerators of convolutional neural networks (CNN). For a given CNN model, hardware accelerator architecture, and FT analysis target, an FPGA-based CNN implementation is generated (with the help of the Tengine framework), and fault injection logic is added. In our first case study, we report how the classification accuracy drop depends on the faults injected into multipliers used in Multiply-and-Accumulate Units of NVDLA inference accelerator executing ResNet-18 CNN. The FT analysis emulated on Zynq UltraScale+ SoC is an order of magnitude faster than software emulation.
\end{abstract}
\vspace{-0.7em}

\section{Introduction}
While convolutional neural networks (CNNs) are inherently error-resilient~\cite{AhmadilivaniTRDJ24}, there is still a critical need for ensuring fault tolerance (FT) in some of their applications, e.g., in autonomous driving and automotive systems. 
This paper deals with FT analysis of hardware inference accelerators of CNNs. 

A common approach to the FT analysis is based on injecting faults into a CNN software model or CNN circuit implementation (depending on the chosen level of abstraction), collecting responses for test stimuli, and computing final statistics~\cite{ahmadilivani2023special}. 
The easiest but the least reliable FT analysis is performed at the CNN execution graph in software (e.g., by introducing ``stuck-at-0'' faults at the outputs of multiplication operations or disconnecting some model's components). More advanced modeling approaches consider the real architecture of the hardware inference accelerator, where a CNN model is executed~\cite{fidelity}. The accelerator is seen as an array of Processing Elements (PEs). The CNN model is executed ``layer by layer'' according to a precomputed execution plan (also called a mapping). 
For example, FIdelity generates a simulation model based on a real accelerator such as the NVDLA (NVIDIA Deep Learning Accelerator) and injects bit flip errors into the software model~\cite{fidelity}.
SAFFIRA employs a simulation model based on a homogeneous PE network and utilizes the Uniform Recurrent Equation system to capture the CNN behavior~\cite{taheri2024saffira}. As this method is computationally expensive, only the most critical CNN layers are  analyzed~\cite{taheri2024saffira}. 
Simulator-based FT analysis often suffers from imprecise modeling of the underlying hardware (i.e., the FT analysis is not reliable) and long execution times~\cite{ahmadilivani2023special,taheri2024saffira,pappalardo:hal-04674751}. 

To overcome these limitations, we propose to extend an existing hardware accelerator to (1) support fault injection (FI) according to user-preferred strategies and (2) perform a fast FT analysis using a hardware emulator implemented on the Zynq UltraScale+ SoC.

Available CNN accelerators developed for FPGAs and ASICs differ in their ability to incorporate additional error injection mechanisms. For instance, the AMD Deep Learning Processor Unit is built entirely from firm IP cores. These cores are configurable units that cannot be modified to emulate faults within specific compute units.
Another example is the DnnWeaver project aiming at converting CNN models into FPGA-compatible formats for computational acceleration~\cite{dnnweaver:micro16}. As this conversion is done through the CNN specification in Caffe, the resulting accelerator is highly specialized to the input CNN. It does not usually show the architecture typical for generic CNN accelerators. Eyeriss~\cite{eyeriss} -- a typical example of a generic accelerator based on a PE array --  has not been designed for FPGAs. 
For purposes of this project, we chose the open-source NVIDIA Deep Learning Accelerator (NVDLA)~\cite{NVDLA:21} utilizing a generic array of PEs. Although primarily intended for ASICs, with some minor modifications, it can be utilized on an FPGAs. 

Our goal is to develop an FPGA-based emulation platform to accelerate fault tolerance analysis of inference accelerators of CNNs.
In addition to accelerating the FT analysis, the platform could be used in the future to validate software FT analysis tools for CNNs and study the impact of introducing various FT mechanisms (such as error-correcting codes or redundant components) into inference accelerators of CNNs.
In the case study, we consider NVDLA accelerator implementing CIFAR-10 image classifier based on ResNet-18 CNN. We analyze (1) how the classification accuracy drop depends on the faults injected into multipliers used in Multiply-and-Accumulate (MAC) units and (2) the speedup of the FT analysis conducted in FPGA contrasted to a software simulation. The proposed emulator is available as open-source software at \textcolor{blue}{\url{https://github.com/ehw-fit/zynq-nvdla-faultinjection}}.

\section{The proposed emulation platform}
\begin{figure}[t]
     \centering\vspace{-1em}
     \includegraphics[width=\linewidth]{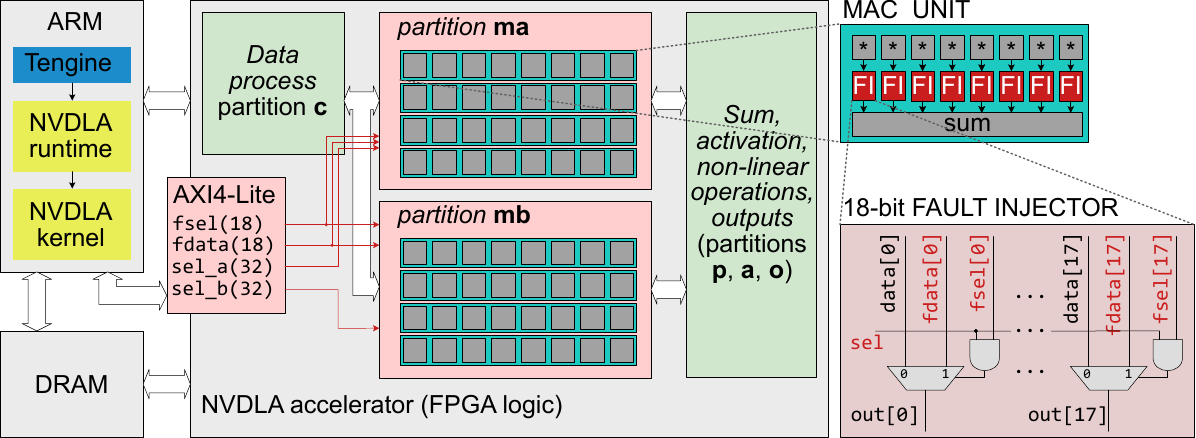}
     \caption{The proposed emulation platform.}\vspace{-1.5em}
     \label{fig:arch}
 \end{figure}

Fig.~\ref{fig:arch} shows the architecture of the proposed emulation platform based on UltraScale+ XCZU7EV Zynq SoC. The accelerator has been created with the help of {NVDLA} Xilinx {FPGA} Mapping tool~\cite{leiwang-nvdla:webpage}. It comprises 8x8 signed 8-bit multipliers arranged into eight MAC units, each of them containing eight multipliers.
The FPGA logic accommodates the array of MAC units (and other operations of the accelerator) whose outputs we equipped with a fault injection logic. A CNN (pre-trained in the Caffe tool) is converted to an execution plan for the accelerator using the Tengine framework~\cite{tengine:webpage}. The execution of the CNN and fault injection are controlled by software running on on-chip ARM cores.
As this is the first experiment, only a small ResNet-18 is considered, utilizing 8-bit precision and excluding features like SRAM cache, reshaping, Winograd convolution, and weight compression. 
Our software modifications of the accelerator implementation~\cite{leiwang-nvdla:webpage} include an improved software stack for the Linux subsystem and an adapted Linux kernel to accommodate the proposed accelerator. 
The fault injection logic controlled via the AXI4 interface bus enables us to override the 18-bit output value of each multiplier with either zero or a constant value, emulating thus a stuck-at error or a pulse fault, respectively. Other fault models can easily be incorporated by modifying the source code.

\section{Experimental results}

Table~\ref{tab:speed} shows the performance and synthesis results for the 8-bit NVDLA accelerator with and without the fault injection support on Zynq, embedded ARM core, and a common PC when executing ResNet-18. The pre-trained ResNet-18 was taken from the Tengine Model ZOO and exhibits the 75.5\% classification accuracy on CIFAR-10.
The hardware implementation performs 4.9$\times$ faster than a single-threat ARM implementation (without FI) and 2.5$\times$ faster than a single-threat AMD Ryzen 7700 implementation. 
When a single global fault has to be injected, the number of LUTs (Look-Up Tables) implementing this fault injection to selected multiplier(s) increases by 18 compared to the NVDLA without the FI support. If the injected value can be dynamically selected, the number of LUTs increases by 0.71\% and the number of flip-flops (FF) by 0.31\%.

\begin{table}[b]
    \centering\vspace*{-1.5em}
    \caption{\small Performance and synthesis results for the 8-bit NVDLA accelerator with and without the fault injection support on Zynq, embedded ARM core, and AMD Ryzen when executing ResNet-18 on CIFAR-10.}\vspace{-2mm}
    \resizebox{\columnwidth}{!}{%
    \begin{tabular}{l c r r r r }\toprule
        Device & Threads / Frequency & Inference (ms) &\#LUT & \#FF  \\\midrule
         ARM Cortex-A53 (Zynq) & 1 / 1.3\,GHz  & 22.68 & --  &  --\\  
         ARM Cortex-A53 (Zynq) & 4 / 1.3\,GHz  & 14.12 & --  &  --\\  
         AMD Ryzen 7 7700  (int8) & 1 / 3.8\,GHz & 11.57 &  -- &  -- \\
        AMD Ryzen 7 7700 (int8) & 4 / 3.8\,GHz &  5.67 &  -- &  -- \\ \midrule
         NVDLA                       & 187.5\,MHz  & 4.59 &   94438 & 104732\\  \midrule
         NVDLA + FI (constant error) &  187.5\,MHz  & 4.59 &  94456 & 104717\\ 
         NVDLA + FI (variable error) & 187.5\,MHz  & 4.59 &  96081 & 106150\\  \bottomrule
    \end{tabular}}
    \label{tab:speed}
\end{table}

\begin{figure}[!htb]
        \centering\vspace{-1em}
	\includegraphics[width=\linewidth]{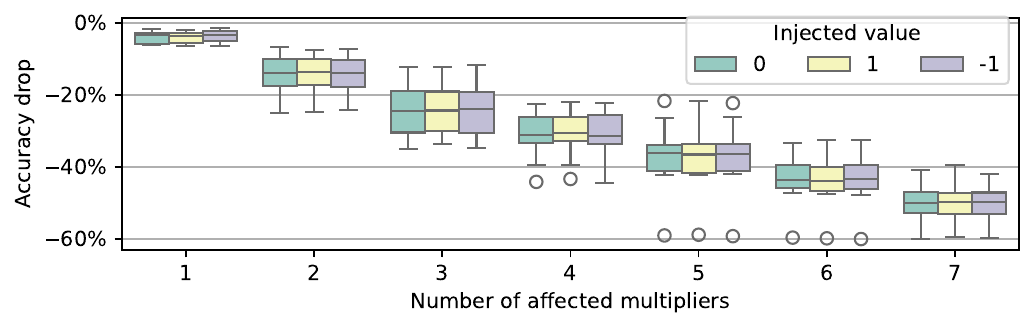}\vspace{-1em}
	\caption{Accuracy drop depending on the number of affected multipliers.}\vspace{-2em}
	\label{fig:multi_cell_mux_count}
\end{figure}

In the first experiment, we injected three types of constant errors (0, 1, and -1) into randomly selected multipliers. The results obtained from 210 FIs are illustrated using box plots in Fig.~\ref{fig:multi_cell_mux_count}. As expected, a noticeable decrease in accuracy is observed when the number of affected multipliers increases, independently of the error value injected.

\begin{figure}[htb!]
	\centering\vspace{-1em}
	\includegraphics[width=\linewidth]{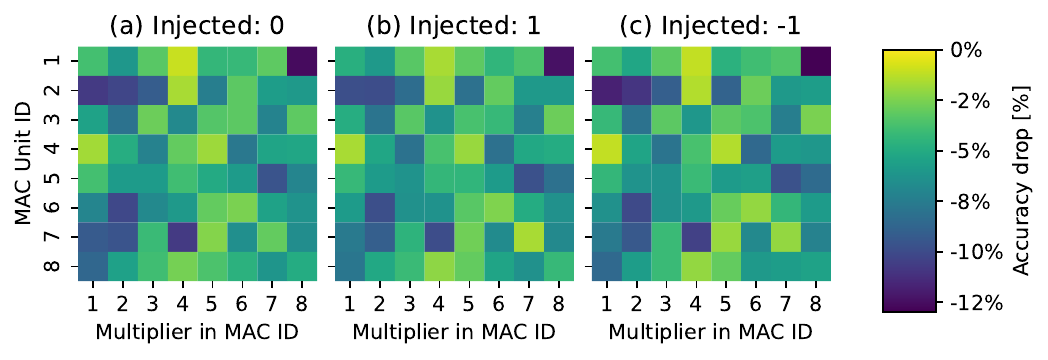}\vspace{-1em}
	\caption{The accuracy drop after injection of (a) 0, (b) 1, and (c) -1 to one multiplier of the MAC unit.}\vspace{-1.8em}
	\label{fig:cell_heatmap}
\end{figure}

In the second experiment, we investigated whether specific positions of the multipliers or the entire MAC unit within the accelerator are more susceptible to faults. One multiplier was always consistently affected, resulting in a complete alteration of the output value from that multiplier. The heat map of the accuracy drop (Fig.~\ref{fig:cell_heatmap}) does not reveal a clear pattern, but it does indicate that some multipliers exhibit greater sensitivity to faults independently of the type of error. The most significant drop in accuracy occurred when the last multiplier of the MAC~1 unit is affected.

\section{Conclusions}
The proposed FPGA-based open-source emulation platform supports real-time fault injection into CNN inference accelerators and its analysis. The area overhead associated with FI is negligible. Compared to recent software approaches, the proposed emulator can perform 217 inferences per second for the whole ResNet-18 while the software engine~\cite{taheri2024saffira} completes 5.8 simulations per second for only two convolutional layers. Our future work will primarily be devoted to improving the emulator in flexibility, configurability, and scalability.

\vspace{0.3em}
\noindent\textit{Acknowledgements:} This work was supported by the Czech Science Foundation project 24-10990S.

\bibliographystyle{IEEEtran}
\bibliography{IEEEabrv,date24}

\end{document}